\documentstyle[12pt,psfig]{article}
\begin{document}

\hfill DUKE-CGTP-2000-14

\hfill hep-th/0008170

\vspace{1.5in}

\begin{center}

{\large\bf Analogues of Discrete Torsion}

{\large\bf for the M-Theory Three-Form }

\vspace{1in}

Eric Sharpe \\
Department of Physics \\
Box 90305 \\
Duke University \\
Durham, NC  27708 \\
{\tt ersharpe@cgtp.duke.edu} \\

 $\,$

\end{center}

In this article we shall outline a derivation of the analogue
of discrete torsion for the M-theory three-form potential.
We find that some of the differences between orbifold
group actions on the $C$ field are classified by $H^3(\Gamma, U(1))$.
We also compute the phases that the low-energy effective action
of a membrane on $T^3$ would see in the analogue of a twisted sector,
and note that they are invariant under the obvious $SL(3, {\bf Z})$ 
action.

\begin{flushleft}
August 2000
\end{flushleft}

\newpage

\tableofcontents

\newpage

\section{Introduction}

Discrete torsion is an obscure-looking degree of freedom
associated with orbifolds.  It was originally discovered
as a set of complex phases that could be added to twisted-sector
contributions to one-loop partition functions \cite{vafa1},
and via the constraint of modular invariance, was found to be
classified by $H^2(\Gamma, U(1))$.  Although a number of papers
have been written on the subject (see for example \cite{vafaed,doug1,doug2}),
it was not until quite recently in \cite{dt1,dt2,dt3} that
a basic geometric understanding of discrete torsion was worked out.

In a nutshell,
\begin{center}
{\it Discrete torsion is the choice of orbifold group action
on the $B$ field.}
\end{center}

More generally, in any theory containing fields with gauge
invariances, to define an orbifold it does not suffice to only
define the orbifold action on the base space.  One must also
specify the orbifold action on the fields.  The choice of orbifold
action on the fields is not unique, precisely because of the possibility
of combining the orbifold group action with a gauge transformation.

For vector fields, these degrees of freedom lead to orbifold Wilson lines.
For $B$ fields, these degrees of freedom lead to discrete torsion.

Put another way, discrete torsion can be understood in purely
mathematical terms, without any reference to string theory -- it
has nothing to do with string theory {\it per se}.

One can now ask, what are the corresponding degrees of freedom for
the other tensor field potentials appearing in string theory?
For example, consider the three-form potential $C$ in $M$ theory.
What degrees of freedom does one have when defining a quotient?

In this paper we describe a preliminary attack on this question.
More precisely, we work out the degrees of freedom describing
possible orbifold group actions on a $C$ field whose curvature
is the image of an element of $H^4({\bf Z})$.
Although this sounds just right to describe the M-theory three-form,
we must caution the reader that we are almost certainly overlooking
important physical subtleties.  For example, we will only consider
the $C$ field in isolation, whereas it is known \cite{fluxquant}
that there are interaction terms which shift the quantization of the
curvature of the $C$ field.  To get a physically correct answer,
we would probably need to take into account that
interaction, which we shall not do.  Also, it has recently been pointed
out \cite{edgreg1,edgreg2} that, in addition to the gravitational
correction just discussed, the M-theory $C$ form potential
should also be understood as the Chern-Simons 3-form on an
$E_8$ gauge bundle with connection.  In order to properly study
orbifold group actions on the M-theory three-form $C$,
one would need to study orbifold group actions on Chern-Simons forms
induced by orbifold group actions on $E_8$ bundles with connection,
which we shall not do.

Despite the fact that the results in this paper may well not be
immediately applicable to the M-theory three-form, we think it is
extremely 
important to make the point that there exists an analogue of
discrete torsion for the M-theory three-form and the other
tensor field potentials of string theory, and that techniques
exist that, in principle, allow one to calculate these degrees of freedom.
 
In section~\ref{orbact}, we study orbifold group actions on 
a $C$ field.  We find that the group $H^3(\Gamma, U(1))$ arises
in describing the differences between some of the orbifold group
actions, just as the group $H^2(\Gamma, U(1))$ arose in \cite{dt1,dt2,dt3}
in describing the difference between some of the orbifold group actions
on the $B$ field.  We also find additional possible orbifold group
actions, beyond those described by $H^3(\Gamma, U(1))$.

In section~\ref{twphase} we work out the analogue of twisted
sector phases for membranes.  Specifically, just as the presence
of the term $\int B$ in the string sigma model leads
to the twisted sector phases that originally characterized
discrete torsion for $B$ fields \cite{dt1,dt3}, 
since membrane low-energy effective actions have a term $\int C$
one expects to have twisted sector phases associated with
membranes. 
We compute these phases for membranes on $T^3$ (described
as an open box on the covering space, with sides identified by
the orbifold group action), and also check
that these phases are invariant under the natural $SL(3, {\bf Z})$
action on $T^3$, the analogue of the modular invariance
constraint on the phases of $B$ field discrete torsion.
(We do not wish to imply that modular invariance is necessarily
a sensible notion for low-energy effective actions on membrane
worldvolumes, but it is certainly amusing to check that the phases
we derive from more fundamental considerations also happen to be
$SL(3, {\bf Z})$-invariant.)

Our methods for understanding analogues of discrete torsion
for the M-theory three-form potential $C$ are purely mathematical.
How would such degrees of freedom show up physically?
If we could compactify M-theory on an additional $S^1$, for example,
how would these degrees  
of freedom be observed in string perturbation theory?
If they could be observed at all in string perturbation
theory, one would surely need to be doing string perturbation
theory in a framework in which Ramond-Ramond fields could
be easily analyzed, as for example \cite{bvw}.  However,
it is not at all clear that such degrees of freedom can be observed
in string perturbation theory.  For example, although
$H^3(\Gamma, U(1))$ enters naturally into ``membrane twisted sectors,''
as we shall see later,
a string worldsheet is one dimension too small to naturally
couple to these degrees of freedom.

In principle one could also calculate analogues of discrete
torsion for the type II Ramond-Ramond tensor field potentials directly.
As these fields are presently believed to be described in terms
of K-theory, one would need a Cheeger-Simons-type description of
K-theory.  Unfortunately, no such description has yet been published,
so we shall not attack the problem of such discrete torsion
analogues here.

The methods in this paper are a direct outgrowth of methods
used in \cite{dt3}, and we shall assume the reader
possesses a working knowledge of that reference.

\section{Orbifold group actions on 2-gerbes}    \label{orbact}

In this section we shall classify orbifold group actions
on three-form potentials, as described in local coordinate patches.
Although we shall often speak of such potentials as
``connections on 2-gerbes,'' in fact we shall not deal at all
with 2-gerbes, but shall merely work in a local coordinate
description.

As described several times earlier, to describe the action of
an orbifold group on a theory containing fields with gauge symmetries,
it is not sufficient merely to describe the action of the orbifold
group on the base space.  In general, one can combine the action
of the orbifold group with gauge transformations, and so one has to
also be careful to specify the action of the orbifold group on
any fields with gauge symmetries.

\subsection{The set of orbifold group actions}

Let $\{ U_{\alpha} \}$ be a ``good invariant'' cover, as 
discussed in \cite{dt1,dt3}.
Then, a three-form field potential is described as a collection
of three-forms $C^{\alpha}$, one for each open set $U_{\alpha}$,
related by gauge transformation on overlaps.
More precisely, 
\begin{eqnarray*}
C^{\alpha} \: - \: C^{\beta} & = & d B^{\alpha \beta} \\
B^{\alpha \beta} \: + \: B^{\beta \gamma} \: + \: B^{\gamma \alpha}
& = & d A^{\alpha \beta \gamma} \\
A^{\beta \gamma \delta} \: - \: A^{\alpha \gamma \delta} 
\: + \: A^{\alpha \beta \delta} \: - \: A^{\alpha \beta \gamma}
& = & d \log h_{\alpha \beta \gamma \delta} \\
\delta h_{\alpha \beta \gamma \delta} & = & 1 
\end{eqnarray*}
(It should be mentioned that the expressions above do not
introduce any new structure not already present in physics.
Readers might object that they are only acquainted with a $C$ field
in physics, and do not recall seeing an associated $B$ or $A$ as above;
however, such readers should be reminded that the $C$ fields
on distinct open patches are related by {\it some} gauge transformation
$B$ -- we are merely making that patch overlap information explicit,
rather than leaving it implicit as is usually done.)
We have implicitly assumed, in writing the above, that
the exterior derivative of the curvature of $C$ vanishes -- that
there are no magnetic sources present -- and that the curvature of $C$
is the image of an element of $H^4({\bf Z})$, i.e., that the curvature
has integral periods.

We shall begin by studying the behavior of the $h_{\alpha \beta \gamma
\delta}$ under pullbacks.  Let $g \in \Gamma$, then
$g^* h_{\alpha \beta \gamma \delta}$ and $h_{\alpha \beta \gamma \delta}$
should differ by only a coboundary:
\begin{equation}    \label{nudefn}
g^* h_{\alpha \beta \gamma \delta} \: = \: \left( h_{\alpha \beta \gamma
\delta} \right) \, \left( \nu^g_{\beta \gamma \delta} \right) \, 
\left( \nu^g_{\alpha \gamma \delta} \right)^{-1} \,
\left( \nu^g_{\alpha \beta \delta} \right) \,
\left( \nu^g_{\alpha \beta \gamma} \right)^{-1}
\end{equation}
If this were not true, one could not even begin to make sense of
an orbifold action, as the $C$ field would not be even remotely symmetric
under the orbifold group action.  Given this assumption, we shall
now follow a self-consistent bootstrap in the spirit of \cite{dt3}
to work out the orbifold group action on the other data defining
the $C$ field.

Now, compare the pullback of $h_{\alpha \beta \gamma \delta}$
by the product $g_1 g_2$ ($g_1, g_2 \in \Gamma$)
and by each group element separately:
\begin{eqnarray*}
(g_1 g_2)^* h_{\alpha \beta \gamma \delta} & = &
\left( h_{\alpha \beta \gamma \delta} \right) \, 
\left( \nu^{g_1 g_2}_{\beta \gamma \delta} \right) \,
\left( \nu^{g_1 g_2}_{\alpha \gamma \delta} \right)^{-1} \cdots \\
g_2^* g_1^* h_{\alpha \beta \gamma \delta} & = &
\left( h_{\alpha \beta \gamma \delta} \right) \,
\left( \nu^{g_2}_{\beta \gamma \delta} \,
g_2^* \nu^{g_1}_{\beta \gamma \delta} \right)
\left( \nu^{g_2}_{\alpha \gamma \delta} \,
g_2^* \nu^{g_1}_{\alpha \gamma \delta} \right)^{-1} \cdots
\end{eqnarray*}
from which we derive\footnote{The attentive reader should,
correctly, object that the equation~(\ref{nuconstr})
is slightly stronger than implied by self-consistency of 
$(g_1 g_2)^* h_{\alpha \beta
\gamma \delta}$ -- that only tells us that the \v{C}ech coboundary of
$\nu^{g_1 g_2}$ must equal the \v{C}ech coboundary of $\nu^{g_2}
g_2^* \nu^{g_1}$.  As in \cite{dt3} we are using self-consistency
to generate constraints of the form of equation~(\ref{nuconstr}) 
on \v{C}ech cochains,
not coboundaries thereof.  If the reader prefers, we are using
self-consistency to generate an ansatz.}
the condition that
$\nu^{g_1 g_2}$ and $\nu^{g_2} g_2^* \nu^{g_1}$ must agree, up
to a coboundary:
\begin{equation}     \label{nuconstr}
\nu^{g_1 g_2}_{\alpha \beta \gamma} \: = \: \left(\nu^{g_2}_{\alpha \beta \gamma} \right)
\left( g_2^* \nu^{g_1}_{\alpha \beta \gamma} \right)
\lambda^{g_1, g_2}_{\alpha \beta} \, \lambda^{g_1, g_2}_{\beta \gamma} \,
\lambda^{g_1, g_2}_{\gamma \alpha}
\end{equation}

Next, we need to examine the coboundaries $\lambda^{g_1, g_2}_{\alpha
\beta}$ more carefully.  There are two distinct ways
to relate $\nu^{g_1 g_2 g_3}$ and $\nu^{g_3} g_3^* \left( \nu^{g_2}
g_2^* \nu^{g_1} \right)$, and they must agree, which constrains
the $\lambda^{g_1, g_2}$.  Specifically,
\begin{eqnarray*}
\nu^{g_1 g_2 g_3}_{\alpha \beta \gamma} & = &
\left( \nu^{g_3}_{\alpha \beta \gamma} \right)
\left( g_3^* \nu^{g_1 g_2}_{\alpha \beta \gamma} \right)
\lambda^{g_1 g_2, g_3}_{\alpha \beta} \,
\lambda^{g_1 g_2, g_3}_{\beta \gamma} \,
\lambda^{g_1 g_2, g_3}_{\gamma \alpha} \\
& = & \left( \nu^{g_3}_{\alpha \beta \gamma} \right)
g_3^* \left( \nu^{g_2}_{\alpha \beta \gamma} \, g_2^* \nu^{g_1}_{\alpha
\beta \gamma} \, \lambda^{g_1, g_2}_{\alpha \beta} \,
\lambda^{g_1, g_2}_{\beta \gamma} \,
\lambda^{g_1, g_2}_{\gamma \alpha} \right)
\lambda^{g_1 g_2, g_3}_{\alpha \beta} \,
\lambda^{g_1 g_2, g_3}_{\beta \gamma} \,
\lambda^{g_1 g_2, g_3}_{\gamma \alpha} \\
\mbox{ also } & = &
\left( \nu^{g_2 g_3}_{\alpha \beta \gamma} \right)
\left( (g_2 g_3)^* \nu^{g_1}_{\alpha \beta \gamma} \right)
\lambda^{g_1, g_2 g_3}_{\alpha \beta} \,
\lambda^{g_1, g_2 g_3}_{\beta \gamma} \,
\lambda^{g_1, g_2 g_3}_{\gamma \alpha} \\
 & = & \left( \nu^{g_3}_{\alpha \beta \gamma} \,
g_3^* \nu^{g_2}_{\alpha \beta \gamma} \,
\lambda^{g_2, g_3}_{\alpha \beta} \,
\lambda^{g_2, g_3}_{\beta \gamma} \,
\lambda^{g_2, g_3}_{\gamma \alpha} \right)
\left( (g_2 g_3)^* \nu^{g_1}_{\alpha \beta \gamma} \right)
\lambda^{g_1, g_2 g_3}_{\alpha \beta} \,
\lambda^{g_1, g_2 g_3}_{\beta \gamma} \,
\lambda^{g_1, g_2 g_3}_{\gamma \alpha} 
\end{eqnarray*}

In order for the equations above to be consistent, we demand
that $\lambda^{g_1 g_2, g_3} g_3^* \lambda^{g_1, g_2}$
and $\lambda^{g_1, g_2 g_3} \lambda^{g_2, g_3}$
agree up to a coboundary:
\begin{equation}    \label{lambdaconstr}
\left( \lambda^{g_1 g_2, g_3}_{\alpha \beta} \right)
\left( g_3^* \lambda^{g_1, g_2}_{\alpha \beta} \right)
\: = \:
\left( \lambda^{g_1, g_2 g_3}_{\alpha \beta} \right)
\left( \lambda^{g_2, g_3}_{\alpha \beta} \right)
\left( \gamma^{g_1, g_2, g_3}_{\alpha} \right)
\left( \gamma^{g_1, g_2, g_3}_{\beta} \right)^{-1}
\end{equation}

Finally, from consistency of the relation~(\ref{lambdaconstr})
in studying products of four group elements, we can derive
a constraint on the coboundaries $\gamma^{g_1, g_2, g_3}$.
Specifically, by solving equation~(\ref{lambdaconstr})
for $(g_3 g_4)^* \lambda^{g_1, g_2}$ in two distinct ways,
one can derive
\begin{equation}    \label{gammaconstr}
\left( \gamma^{g_1, g_2, g_3 g_4}_{\alpha} \right) \, 
\left( \gamma^{g_1 g_2, g_3, g_4}_{\alpha} \right) \: = \:
\left( \gamma^{g_1, g_2 g_3, g_4}_{\alpha} \right) \,
\left( \gamma^{g_2, g_3, g_4}_{\alpha} \right) \,
\left( g_4^* \gamma^{g_1, g_2, g_3}_{\alpha} \right)
\end{equation}

So far we have derived the form of a lift of $\Gamma$ to merely
the \v{C}ech cocycles $h_{\alpha \beta \gamma \delta}$;
we still need to describe how the orbifold group acts on 
the local three-forms, two-forms, and one-forms defining the connection
on the 2-gerbe.

We can write
\begin{eqnarray*}
g^* C^{\alpha} & = & C^{\alpha} \: + \: C(g)^{\alpha} \\
g^* B^{\alpha \beta} & = & B^{\alpha \beta} \: + \: B(g)^{\alpha \beta} \\
g^* A^{\alpha \beta \gamma} & = & A^{\alpha \beta \gamma} \: + \:
A(g)^{\alpha \beta \gamma}
\end{eqnarray*}
for some local three-forms $C(g)^{\alpha}$, two-forms
$B(g)^{\alpha \beta}$, and one-forms $A(g)^{\alpha \beta \gamma}$.
Using self-consistency, we shall derive more meaningful expressions
for $C(g)^{\alpha}$, $B(g)^{\alpha \beta}$, and $A(g)^{\alpha \beta \gamma}$.

Expand both sides of the expression
\begin{displaymath}
g^* \left( \, C^{\alpha} \: - \: C^{\beta} \, \right) \: = \:
g^* d B^{\alpha \beta}
\end{displaymath}
to show that
\begin{equation}  \label{cg1}
C(g)^{\alpha} \: - \: C(g)^{\beta} \: = \: d B(g)^{\alpha \beta \gamma}
\end{equation}
Similarly, use
\begin{displaymath}
g^* \left( \, B^{\alpha \beta} \: + \: B^{\beta \gamma} \: + \:
B^{\gamma \alpha} \, \right) \: = \: d A^{\alpha \beta \gamma}
\end{displaymath}
to show
\begin{equation}    \label{bg1}
B(g)^{\alpha \beta} \: + \: B(g)^{\beta \gamma} \: + \:
B(g)^{\gamma \alpha} \: = \: d A(g)^{\alpha \beta \gamma}
\end{equation}
and use
\begin{displaymath}
g^* \left( \, \delta A^{\alpha \beta \gamma} \, \right) \: = \:
g^* d \log h_{\alpha \beta \gamma \delta}
\end{displaymath}
to show
\begin{equation}
A(g)^{\alpha \beta \gamma} \: = \: d \log \nu^g_{\alpha \beta \gamma}
\: + \: \Lambda^{(1)}(g)^{\alpha \beta} \: + \:
\Lambda^{(1)}(g)^{\beta \gamma} \: + \:
\Lambda^{(1)}(g)^{\gamma \alpha}
\end{equation}
for some local one-forms $\Lambda^{(1)}(g)^{\alpha \beta}$.
Plugging this expression back into equation~(\ref{bg1}),
we find
\begin{equation}
B(g)^{\alpha \beta} \: = \: d \Lambda^{(1)}(g)^{\alpha \beta} \: + \:
\Lambda^{(2)}(g)^{\alpha} \: - \:
\Lambda^{(2)}(g)^{\beta}
\end{equation}
for some local two-forms $\Lambda^{(2)}(g)^{\alpha}$.
Plugging this back into equation~(\ref{cg1}),
we find
\begin{equation}
C(g)^{\alpha} \: = \: d \Lambda^{(2)}(g)^{\alpha}
\end{equation}

To summarize progress so far, we have found
\begin{eqnarray*}
g^* C^{\alpha} & = & C^{\alpha} \: + \: d \Lambda^{(2)}(g)^{\alpha} \\
g^* B^{\alpha \beta} & = & B^{\alpha \beta} \: + \:
d \Lambda^{(1)}(g)^{\alpha \beta} \: + \: \Lambda^{(2)}(g)^{\alpha}
\: - \: \Lambda^{(2)}(g)^{\beta} \\
g^* A^{\alpha \beta \gamma} & = & A^{\alpha \beta \gamma} \: + \:
d \log \nu^g_{\alpha \beta \gamma} \: + \:
\Lambda^{(1)}(g)^{\alpha \beta} \: + \:
\Lambda^{(1)}(g)^{\beta \gamma} \: + \:
\Lambda^{(1)}(g)^{\gamma \alpha}
\end{eqnarray*}
for some forms $\Lambda^{(1)}(g)^{\alpha \beta}$ and 
$\Lambda^{(2)}(g)^{\alpha}$.

Next, we need to determine how $\Lambda^{(1)}(g_1 g_2)^{\alpha \beta}$
and $\Lambda^{(2)}(g_1 g_2)^{\alpha}$ are related to
$\Lambda^{(1)}(g_1)^{\alpha \beta}$, $\Lambda^{(1)}(g_2)^{\alpha \beta}$,
$\Lambda^{(2)}(g_1)^{\alpha}$, and $\Lambda^{(2)}(g_2)^{\alpha}$.

By evaluating $(g_1 g_2)^* C^{\alpha}$ in two different ways,
we find
\begin{equation}  \label{lambda2}
\Lambda^{(2)}(g_1 g_2)^{\alpha} \: = \:
\Lambda^{(2)}(g_2)^{\alpha} \: + \: g_2^* \Lambda^{(2)}(g_1)^{\alpha}
\: + \: d \Lambda^{(3)}(g_1, g_2)^{\alpha}
\end{equation}
for some local one-forms $\Lambda^{(3)}(g_1, g_2)^{\alpha}$.

By evaluating $(g_1 g_2)^* A^{\alpha \beta \gamma}$ in two different
ways, we find
\begin{displaymath}
\delta \left( \, d \log \lambda^{g_1, g_2}_{\alpha \beta} \: + \:
\Lambda^{(1)}(g_1 g_2)^{\alpha \beta} \, \right)
\: = \:
\delta \left( \, \Lambda^{(1)}(g_2)^{\alpha \beta} \: + \:
g_2^* \Lambda^{(1)}(g_1)^{\alpha \beta} \, \right)
\end{displaymath}
By evaluating $(g_1 g_2)^* B^{\alpha \beta \gamma}$ in two different
ways, we find
\begin{displaymath}
d \left( \, \Lambda^{(1)}(g_1 g_2)^{\alpha \beta} \: + \:
\Lambda^{(3)}(g_1, g_2)^{\alpha} \: - \: \Lambda^{(3)}(g_1, g_2)^{\beta}
\, \right)
\: = \: d \left( \,
\Lambda^{(1)}(g_2)^{\alpha \beta} \: + \: g_2^* \Lambda^{(1)}(g_1)^{\alpha
\beta} \, \right)
\end{displaymath}
After combining these two equations, we find
\begin{equation}   \label{lambda1}
\Lambda^{(1)}(g_1 g_2)^{\alpha \beta} \: + \:
\Lambda^{(3)}(g_1, g_2)^{\alpha} \: - \:
\Lambda^{(3)}(g_1, g_2)^{\beta} \: + \: d \log \lambda^{g_1, g_2}_{\alpha 
\beta}
\: = \:
\Lambda^{(1)}(g_2)^{\alpha \beta} \: + \:
g_2^* \Lambda^{(1)}(g_1)^{\alpha \beta}
\end{equation}

By using equation~(\ref{lambda2}) to evaluate
$\Lambda^{(2)}(g_1 g_2 g_3)^{\alpha}$ in two different ways, we find
\begin{displaymath}
d \left( \, \Lambda^{(3)}(g_2, g_3)^{\alpha} \: + \:
\Lambda^{(3)}(g_1, g_2 g_3)^{\alpha} \, \right)
\: = \: d \left( \,
g_3^* \Lambda^{(3)}(g_1, g_2)^{\alpha} \: + \:
\Lambda^{(3)}(g_1 g_2, g_3)^{\alpha} \,
\right)
\end{displaymath}
By using equation~(\ref{lambda1}) to evaluate
$\Lambda^{(1)}(g_1 g_2 g_3)^{\alpha \beta}$ in two different ways,
we find
\begin{displaymath}
\delta \left( \, \Lambda^{(3)}(g_2, g_3)^{\alpha} \: + \:
\Lambda^{(3)}(g_1, g_2 g_3)^{\alpha} \, \right)
\: = \:
\delta \left( \, g_3^* \Lambda^{(3)}(g_1, g_2)^{\alpha}
\: + \: \Lambda^{(3)}(g_1 g_2, g_3)^{\alpha} 
\: + \: d \log \gamma_{\alpha}^{g_1, g_2, g_3} \,  \right)
\end{displaymath}
Combining these two equations, we find
\begin{equation}    \label{Lambda3gam}
\Lambda^{(3)}(g_2, g_3)^{\alpha} \: + \:
\Lambda^{(3)}(g_1, g_2 g_3)^{\alpha} \: = \:
g_3^* \Lambda^{(3)}(g_1, g_2)^{\alpha} \: + \:
\Lambda^{(3)}(g_1 g_2, g_3)^{\alpha}
\: + \: d \log \gamma_{\alpha}^{g_1, g_2, g_3}
\end{equation}
(In principle, we could add a term of the form
\begin{displaymath}
d \Lambda^{(4)}(g_1, g_2, g_3)^{\alpha} \: - \:
d \Lambda^{(4)}(g_1, g_2, g_3)^{\beta}
\end{displaymath}
to one side of the expression above, i.e., something annihilated by
both $d$ and $\delta$; however, the only quantities that we shall use
are $d \Lambda^{(3)}$ and $\delta \Lambda^{(3)}$, so we shall
ignore such additional possible terms.)

The calculation above completes our derivation of how
orbifold groups act on 2-gerbes with connection,
at the level of \v{C}ech cocycles.  We have included a table
below to summarize the results:

\begin{eqnarray*}
g^* C^{\alpha} & = & C^{\alpha} \: + \: d \Lambda^{(2)}(g)^{\alpha} \\
g^* B^{\alpha \beta} & = & B^{\alpha \beta} \: + \:
d \Lambda^{(1)}(g)^{\alpha \beta} \: + \:
\Lambda^{(2)}(g)^{\alpha} \: - \: \Lambda^{(2)}(g)^{\beta} \\
g^* A^{\alpha \beta \gamma} & = & A^{\alpha \beta \gamma} \: + \:
d \log \nu^g_{\alpha \beta \gamma} \: + \: \Lambda^{(1)}(g)^{\alpha \beta}
\: + \: \Lambda^{(1)}(g)^{\beta \gamma} \: + \:
\Lambda^{(1)}(g)^{\gamma \alpha} \\
g^* h_{\alpha \beta \gamma \delta} & = &
\left( h_{\alpha \beta \gamma \delta} \right) \, 
\left( \nu^g_{\beta \gamma \delta} \right) \,
\left( \nu^g_{\alpha \gamma \delta} \right)^{-1} \,  
\left( \nu^g_{\alpha \beta \delta} \right) \,
\left( \nu^g_{\alpha \beta \gamma} \right)^{-1} \\
\, & \, & \, \\
\Lambda^{(2)}(g_1 g_2)^{\alpha} & = &
\Lambda^{(2)}(g_2)^{\alpha} \: + \: g_2^* \Lambda^{(2)}(g_1)^{\alpha} 
\: + \: d \Lambda^{(3)}(g_1, g_2)^{\alpha} \\
\Lambda^{(1)}(g_1 g_2)^{\alpha \beta} 
& = &
\Lambda^{(1)}(g_2)^{\alpha \beta} \: + \: g_2^* \Lambda^{(1)}(g_1)^{\alpha
\beta}  \: - \: \Lambda^{(3)}(g_1, g_2)^{\alpha} \: + \:
\Lambda^{(3)}(g_1, g_2)^{\beta} \\
 & & \makebox[50pt][r]{$\,$} \: - \:  
d \log \lambda^{g_1, g_2}_{\alpha \beta} \\
\Lambda^{(3)}(g_2, g_3)^{\alpha} \: + \: \Lambda^{(3)}(g_1, g_2 g_3)^{\alpha}
& = & g_3^* \Lambda^{(3)}(g_1, g_2)^{\alpha} \: + \:
\Lambda^{(3)}(g_1 g_2, g_3)^{\alpha} \: + \: d \log \gamma^{g_1, g_2, g_3}_{
\alpha} \\
\, & \, & \, \\
\nu^{g_1 g_2}_{\alpha \beta \gamma} & = &
\left( \nu^{g_2}_{\alpha \beta \gamma} \right) \, \left(
g_2^* \nu^{g_1}_{\alpha \beta \gamma} \right) \,
\left( \lambda^{g_1, g_2}_{\alpha \beta} \right) \, \left(
\lambda^{g_1, g_2}_{\beta \gamma} \right) \,
\left( \lambda^{g_1, g_2}_{\gamma \alpha} \right) \\
\left( \lambda^{g_1 g_2, g_3}_{\alpha \beta} \right) 
\, \left( g_3^* \lambda^{g_1, g_2}_{\alpha
\beta} \right) & = & \left( \lambda^{g_1, g_2 g_3}_{\alpha \beta} \right) \,
\left( \lambda^{g_2, g_3}_{\alpha \beta} \right) 
\, \left( \gamma^{g_1, g_2, g_3}_{\alpha}
\right) \, \left( \gamma^{g_1, g_2, g_3}_{\beta} \right)^{-1} \\
\left( \gamma^{g_1, g_2, g_3 g_4}_{\alpha} \right) \,
\left( \gamma^{g_1 g_2, g_3, g_4}_{\alpha} \right)
& = &
\left( \gamma^{g_1, g_2 g_3, g_4}_{\alpha} \right) \,
\left( \gamma^{g_2, g_3, g_4}_{\alpha} \right) \,
\left( g_4^* \gamma^{g_1, g_2, g_3}_{\alpha} \right)
\end{eqnarray*}
where $\nu^g_{\alpha \beta \gamma}$, $\lambda^{g_1, g_2}_{\alpha \beta}$,
$\gamma^{g_1, g_2, g_3}_{\alpha}$, $\Lambda^{(1)}(g)^{\alpha \beta}$,
$\Lambda^{(2)}(g)^{\alpha}$, and $\Lambda^{(3)}(g_1, g_2)^{\alpha}$
are structures introduced to define the orbifold group action.

\subsection{Differences between orbifold group actions}

In describing orbifold $U(1)$ Wilson lines, the group
$H^1(\Gamma, U(1))$ arises as differences between orbifold group actions;
similarly, in describing discrete torsion, $H^2(\Gamma, U(1))$ arises
in describing the differences between orbifold group actions.
Here, we shall also study differences between orbifold group actions,
and, at the end of the day, we shall discover that some of the differences
between orbifold group actions are classified by $H^3(\Gamma, U(1))$.

Let one lift of the orbifold group $\Gamma$ be denoted using the same
notation as in the last section, and let a second lift be denoted
by overlining.

To begin, define 
\begin{displaymath}
\Upsilon^{g}_{\alpha \beta \gamma} \: = \:
\frac{ \nu^g_{\alpha \beta \gamma} }{ \overline{\nu}^g_{\alpha \beta
\gamma} }
\end{displaymath}
From equation~(\ref{nudefn}), as applied to each of the two lifts,
it is straightforward to derive that the $\Upsilon$ are \v{C}ech
cocycles:
\begin{equation}    \label{upsbas}
\delta \Upsilon^g \: = \: 1
\end{equation}
In fact, such \v{C}ech cocycles define a 1-gerbe.
This should not be surprising -- after all, a gauge transformation of a 1-gerbe
is a 0-gerbe, i.e., a principal $U(1)$ bundle.  Here, we see
that the difference between two lifts of orbifold group actions to
2-gerbes is a gauge transformation of the 2-gerbe, namely,
a set of 1-gerbes.

Define 
\begin{eqnarray*}
{\cal B}(g)^{\alpha} & = & \Lambda^{(2)}(g)^{\alpha} \: - \:
\overline{\Lambda}^{(2)}(g)^{\alpha} \\
{\cal A}(g)^{\alpha \beta} & = & \overline{\Lambda}^{(1)}(g)^{\alpha \beta}
\: - \: \Lambda^{(1)}(g)^{\alpha \beta} 
\end{eqnarray*}
By comparing $g^* A^{\alpha \beta \gamma}$ expressed in terms of the
two different lifts, we find that
\begin{equation}  \label{atoup}
{\cal A}(g)^{\alpha \beta} \: + \:
{\cal A}(g)^{\beta \gamma} \: + \:
{\cal A}(g)^{\gamma \alpha} \: = \:
d \log \Upsilon^g_{\alpha \beta \gamma}
\end{equation}
and by comparing $g^* B^{\alpha \beta}$ expressed in terms of the
two different lifts, we find that
\begin{equation}  \label{btoa}
{\cal B}(g)^{\alpha} \: - \: {\cal B}(g)^{\beta} \: = \:
d {\cal A}(g)^{\alpha \beta}
\end{equation}

From equations~(\ref{btoa}), (\ref{atoup}), and (\ref{upsbas})
we see that the data $( {\cal B}(g)^{\alpha}, {\cal A}(g)^{\alpha \beta},
\Upsilon^g_{\alpha \beta \gamma})$ defines a 1-gerbe with connection,
in the language of \cite{hitchin,cjthes}.  Again, as mentioned a
few paragraphs ago, a 1-gerbe with connection defines a gauge
transformation, so we have just learned that (part of) the difference
between any two lifts of the orbifold group action is a set of
gauge transformations, one for each element of the orbifold group.
This is precisely analogous to orbifold $U(1)$ Wilson lines,
where we observed that any two lifts of the action of the orbifold group
differ by a set of gauge transformations.

Furthermore, from comparing $g^* C^{\alpha}$ expressed in terms of the
two different lifts, we find that
\begin{equation}
d {\cal B}(g)^{\alpha} \: = \: 0
\end{equation}
In other words, to preserve the connection on the 2-gerbe,
the possible gauge transformations are restricted to 1-gerbes
with flat connection, just as for orbifold $U(1)$ Wilson lines,
to preserve the connection, gauge transformations were restricted to
be constant.

Of course, there is additional structure present here.
Define
\begin{displaymath}
\Omega^{g_1, g_2}_{\alpha \beta} \: = \: \frac{ 
\lambda^{g_1, g_2}_{\alpha \beta} }{
\overline{\lambda}^{g_1, g_2}_{\alpha \beta} }
\end{displaymath}
From dividing the two expressions
\begin{eqnarray*}
\nu^{g_1 g_2}_{\alpha \beta \gamma} & = &
\left( \nu^{g_2}_{\alpha \beta \gamma} \right) \, \left(
g_2^* \nu^{g_1}_{\alpha \beta \gamma} \right) \,
\left( \lambda^{g_1, g_2}_{\alpha \beta} \right) \, \left(
\lambda^{g_1, g_2}_{\beta \gamma} \right) \,
\left( \lambda^{g_1, g_2}_{\gamma \alpha} \right) \\
\overline{\nu}^{g_1 g_2}_{\alpha \beta \gamma} & = &
\left( \overline{\nu}^{g_2}_{\alpha \beta \gamma} \right) \, \left(
g_2^* \overline{\nu}^{g_1}_{\alpha \beta \gamma} \right) \,
\left( \overline{\lambda}^{g_1, g_2}_{\alpha \beta} \right) \, \left(
\overline{\lambda}^{g_1, g_2}_{\beta \gamma} \right) \,
\left( \overline{\lambda}^{g_1, g_2}_{\gamma \alpha} \right) 
\end{eqnarray*}
we see that
\begin{equation}    \label{Omasmap}
\Upsilon^{g_1 g_2}_{\alpha \beta \gamma} \: = \:
\left( \Upsilon^{g_2}_{\alpha \beta \gamma} \right) \,
\left( g_2^* \Upsilon^{g_1}_{\alpha \beta \gamma} \right) \,
\left( \Omega^{g_1, g_2}_{\alpha \beta} \right) \,
\left( \Omega^{g_1, g_2}_{\beta \gamma} \right) \,
\left( \Omega^{g_1, g_2}_{\gamma \alpha} \right)
\end{equation}
which is the statement in local trivializations
\cite{dt2} that
$\Omega^{g_1, g_2}$ defines a
map between 1-gerbes.
(We should be slightly careful -- although the $\Omega^{g_1, g_2}$ define
a map between 1-gerbes, they are not sufficient to completely
define a map between 1-gerbes with connection.  In a few paragraphs
we will find the additional structure needed to complete the 
maps $\Omega^{g_1, g_2}$ to define a morphism of 1-gerbes with connection.)
Denoting the 1-gerbes described by the cocycles $\Upsilon^g_{\alpha
\beta \gamma}$ by $\Upsilon^g$, we can describe the maps $\Omega^{g_1, g_2}$
defined by the \v{C}ech cochains $\Omega^{g_1, g_2}_{\alpha \beta}$
as maps
\begin{displaymath}
\Omega^{g_1, g_2}: \: \Upsilon^{g_2} \otimes g_2^* \Upsilon^{g_1}
\: \longrightarrow \: \Upsilon^{g_1 g_2}
\end{displaymath}

We should take a moment to explain the meaning of $\otimes$ for
1-gerbes.  A technical definition is given in
\cite[section 4.1]{brylinski}.  However, for the purposes of
most readers, it should suffice to think of $\otimes$ as
an operation which associates a 1-gerbe to \v{C}ech cocycles
given by the product of \v{C}ech cocycles defining two other
1-gerbes.

The constraint~(\ref{lambdaconstr}) becomes the constraint
that there is a well-defined relationship between 
$\Upsilon^{g_1 g_2 g_3}$ and $\Upsilon^{g_3} \otimes
g_3^* \left( \Upsilon^{g_2} \otimes g_2^* \Upsilon^{g_1} \right)$.
As we shall explain more concretely in a moment, 
constraint~(\ref{lambdaconstr}) implies that
the following diagram\footnote{Experts will recognize
that we are being slightly sloppy, in that, for example,
we identify $(g_2 g_3)^* \Upsilon^{g_1}$ with $g_3^* g_2^* \Upsilon^{g_1}$.
Strictly speaking, for 1-gerbes the story is slightly more complicated,
in that they are merely canonically isomorphic.
However, at the level of \v{C}ech cocycles, such technical
distinctions are not present, so we omit them from the discussion.} 
commutes, up to isomorphism
of maps:
\begin{equation}   \label{omegacomm}
\begin{array}{ccc}
\Upsilon^{g_3} \otimes g_3^* \left( \Upsilon^{g_2} \otimes
g_2^* \Upsilon^{g_1} \right) & \stackrel{
g_3^* \Omega^{g_1, g_2} }{\longrightarrow} &
\Upsilon^{g_3} \otimes g_3^* \Upsilon^{g_1 g_2} \\
\makebox[0pt][r]{$\scriptstyle{\Omega^{g_2, g_3}}$} \downarrow & &
\downarrow \makebox[0pt][l]{$\scriptstyle{\Omega^{g_1 g_2, g_3}}$} \\
\Upsilon^{g_2 g_3} \otimes (g_2 g_3)^* \Upsilon^{g_1} &
\stackrel{ \Omega^{g_1, g_2 g_3} }{\longrightarrow } &
\Upsilon^{g_1 g_2 g_3} 
\end{array}
\end{equation}
Again, the correct statement is that the diagram above commutes
up to isomorphism of maps:
\begin{displaymath}
\Omega^{g_1 g_2, g_3} \circ g_3^* \Omega^{g_1, g_2} \:
\cong \: \Omega^{g_1, g_2 g_3} \circ \Omega^{g_2, g_3}
\end{displaymath}
To understand why these maps are merely isomorphic -- or what
a map between maps is, in this context -- recall
from \cite{dt2} that 1-gerbes can be understood as sheaves of
categories, roughly speaking, and maps between them are
sheaves of functors.  A ``map between maps,'' in this context,
is a sheaf of natural transformations.

More concretely, these ``maps between maps'' are defined by 
the \v{C}ech cochains $\gamma^{g_1, g_2, g_3}$.
Define
\begin{displaymath}
\omega^{g_1, g_2, g_3}_{\alpha} \: = \:
\frac{  \gamma^{g_1, g_2, g_3}_{\alpha} }{
\overline{\gamma}^{g_1, g_2, g_3}_{\alpha} }
\end{displaymath}
From the analogues of equation~(\ref{lambdaconstr}) for the two
orbifold group actions, 
we see that
each \v{C}ech cochain $\omega^{g_1, g_2, g_3}_{\alpha}$ obeys
\begin{equation}
\left( \Omega^{g_1 g_2, g_3}_{\alpha \beta} \right) \,
\left( g_3^* \Omega^{g_1, g_2}_{\alpha \beta} \right) \: = \:
\left( \Omega^{g_1, g_2 g_3}_{\alpha \beta} \right) \,
\left( \Omega^{g_2, g_3}_{\alpha \beta} \right) \,
\left( \omega^{g_1, g_2, g_3}_{\alpha} \right) \,
\left( \omega^{g_1, g_2, g_3}_{\beta} \right)^{-1}
\end{equation}
This means that the cochains $\omega^{g_1, g_2, g_3}_{\alpha}$
define a map $\omega^{g_1, g_2, g_3}$ between maps:
\begin{displaymath}
\omega^{g_1, g_2, g_3}: \: \Omega^{g_1, g_2 g_3} \circ \Omega^{g_2, g_3}
\: \longrightarrow \: \Omega^{g_1 g_2, g_3} \circ
g_3^* \Omega^{g_1, g_2}
\end{displaymath}

From equation~(\ref{gammaconstr}), we find that the $\omega$ are constrained
as
\begin{equation}      \label{xiconstr}
\left( \omega^{g_1, g_2, g_3 g_4}_{\alpha} \right) \, 
\left( \omega^{g_1 g_2, g_3, g_4}_{\alpha}  \right) \: = \:
\left( \omega^{g_1, g_2 g_3, g_4}_{\alpha} \right) \,
\left( \omega^{g_2, g_3, g_4}_{\alpha} \right) \,
\left( g_4^* 
\omega^{g_1, g_2, g_3}_{\alpha} \right)
\end{equation}
As a consistency check, the attentive reader will note that both
$\omega^{g_1 g_2, g_3, g_4} \circ \omega^{g_1, g_2, g_3 g_4}$
and $g_4^* \omega^{g_1, g_2, g_3} \circ \omega^{g_1, g_2 g_3, g_4} 
\circ \omega^{g_2, g_3, g_4}$
map
\begin{displaymath}
\Omega^{g_1, g_2 g_3 g_4} \circ \Omega^{g_2, g_3 g_4} \circ
\Omega^{g_3, g_4}
\: \longrightarrow \:
\Omega^{g_1 g_2 g_3, g_4} \circ g_4^* \Omega^{g_1 g_2, g_3} \circ
(g_3 g_4)^* \Omega^{g_1, g_2} 
\end{displaymath}
From equation~(\ref{xiconstr}), we see that the two possible
maps are the same.

Finally, define
\begin{equation}
\theta(g_1, g_2)^{\alpha} \: = \: \overline{\Lambda}^{(3)}(g_1, g_2)^{\alpha}
\: - \: \Lambda^{(3)}(g_1, g_2)^{\alpha}
\end{equation}
We shall see concretely in a moment that $\theta(g_1, g_2)$ combines
with $\Omega^{g_1, g_2}$ to define a map of 1-gerbes with connection
(whereas $\Omega^{g_1, g_2}$ by itself merely defined a map of 1-gerbes,
and was not sufficient to describe the action on the connection).

From equation~(\ref{lambda2}) for each orbifold group action,
we immediately see that
\begin{equation}
{\cal B}(g_1 g_2)^{\alpha} \: = \:
{\cal B}(g_2)^{\alpha} \: + \:
g_2^* {\cal B}(g_1)^{\alpha} \: - \:
d \theta(g_1, g_2)^{\alpha}
\end{equation}
From equation~(\ref{lambda1}) for each orbifold group action,
we find that
\begin{equation}     \label{calAthetaOm}
{\cal A}(g_1 g_2)^{\alpha \beta} \: = \:
{\cal A}(g_2)^{\alpha \beta} \: + \: g_2^* {\cal A}(g_1)^{\alpha \beta}
\: - \: \theta(g_1, g_2)^{\alpha} \: + \: \theta(g_1, g_2)^{\beta}
\: + \: d \log \Omega^{g_1, g_2}_{\alpha \beta}
\end{equation}
The two equations above mean that the pair $\left(\Omega^{g_1, g_2},
\theta(g_1, g_2) \right)$ define a map of 1-gerbes with connection:
\begin{displaymath}
%\left( \Omega^{g_1, g_2}, \theta(g_1, g_2) \right):
\left( {\cal B}(g_2)^{\alpha}, {\cal A}(g_2)^{\alpha \beta},
\Upsilon^{g_2}_{\alpha \beta \gamma} \right) 
\otimes g_2^* \left(
{\cal B}(g_1)^{\alpha}, {\cal A}(g_1)^{\alpha \beta},
\Upsilon^{g_1}_{\alpha \beta \gamma} \right) \: \longrightarrow \:
\left( {\cal B}(g_1 g_2)^{\alpha}, {\cal A}(g_1 g_2)^{\alpha \beta},
\Upsilon^{g_1 g_2}_{\alpha \beta \gamma} \right)
\end{displaymath}

From equation~(\ref{Lambda3gam}) for each orbifold group action, 
we find that
\begin{equation}   \label{thetaom}
\theta(g_2, g_3)^{\alpha} \: + \: \theta(g_1, g_2 g_3)^{\alpha} \: = \:
g_3^* \theta(g_1, g_2)^{\alpha} \: + \:
\theta(g_1 g_2, g_3)^{\alpha} \: - \: d \log \omega^{g_1, g_2, g_3}_{\alpha}
\end{equation} 
which means that diagram~(\ref{omegacomm}) commutes up to
isomorphism of maps when interpreted as a diagram of 1-gerbes
with connection, not merely gerbes.  Put another way,
this means that the ``map between maps'' $\omega^{g_1, g_2, g_3}$
is not only a map between the morphisms of 1-gerbes
$\Omega^{g_1, g_2}$, but is also a map between the morphisms
of 1-gerbes with connection $\left(
\Omega^{g_1, g_2}, \theta(g_1, g_2) \right)$.

To summarize, we have found that the difference between any
two orbifold group action on a set of $C$ fields is defined
by a set of 1-gerbes with flat connection $({\cal B}(g)^{\alpha},
{\cal A}(g)^{\alpha \beta}, \Upsilon^g_{\alpha \beta \gamma})$,
one for each $g \in \Gamma$, plus maps $\left(\Omega^{g_1, g_2},
\theta(g_1, g_2) \right)$ between 1-gerbes with connection
\begin{displaymath}
%\left( \Omega^{g_1, g_2}, \theta(g_1, g_2) \right):
\left( {\cal B}(g_2)^{\alpha}, {\cal A}(g_2)^{\alpha \beta},
\Upsilon^{g_2}_{\alpha \beta \gamma} \right)
\otimes g_2^* \left(
{\cal B}(g_1)^{\alpha}, {\cal A}(g_1)^{\alpha \beta},
\Upsilon^{g_1}_{\alpha \beta \gamma} \right) \: \longrightarrow \:
\left( {\cal B}(g_1 g_2)^{\alpha}, {\cal A}(g_1 g_2)^{\alpha \beta},
\Upsilon^{g_1 g_2}_{\alpha \beta \gamma} \right)
\end{displaymath}
and ``maps between maps'' $\omega^{g_1, g_2, g_3}$:
\begin{displaymath}
\omega^{g_1, g_2, g_3}: \: \Omega^{g_1, g_2 g_3} \circ \Omega^{g_2, g_3}
\: \longrightarrow \: \Omega^{g_1 g_2, g_3} \circ
g_3^* \Omega^{g_1, g_2}
\end{displaymath}
The 1-gerbe morphisms $\left(
\Omega^{g_1, g_2}, \theta(g_1, g_2) \right)$ are constrained
to make the following diagram commute
\begin{displaymath}   
\begin{array}{ccc}
\Upsilon^{g_3} \otimes g_3^* \left( \Upsilon^{g_2} \otimes
g_2^* \Upsilon^{g_1} \right) & \stackrel{
g_3^* \Omega^{g_1, g_2} }{\longrightarrow} &
\Upsilon^{g_3} \otimes g_3^* \Upsilon^{g_1 g_2} \\
\makebox[0pt][r]{$\scriptstyle{\Omega^{g_2, g_3}}$} \downarrow & &
\downarrow \makebox[0pt][l]{$\scriptstyle{\Omega^{g_1 g_2, g_3}}$} \\
\Upsilon^{g_2 g_3} \otimes (g_2 g_3)^* \Upsilon^{g_1} &
\stackrel{ \Omega^{g_1, g_2 g_3} }{\longrightarrow } &
\Upsilon^{g_1 g_2 g_3}
\end{array}
\end{displaymath}
up to isomorphisms of maps defined by $\omega^{g_1, g_2, g_3}$,
%\begin{displaymath}
%\omega^{g_1, g_2, g_3}: \: \Omega^{g_1, g_2 g_3} \circ \Omega^{g_2, g_3}
%\: \longrightarrow \: \Omega^{g_1 g_2, g_3} \circ
%g_3^* \Omega^{g_1, g_2}
%\end{displaymath}
and the maps of maps $\omega^{g_1, g_2, g_3}$ are constrained to obey
\begin{displaymath}
\omega^{g_1 g_2, g_3, g_4} \circ \omega^{g_1, g_2, g_3 g_4} \: = \:
g_4^* \omega^{g_1, g_2, g_3} \circ \omega^{g_1, g_2 g_3, g_4} \circ
\omega^{g_2, g_3, g_4}
\end{displaymath}

\subsection{Residual gauge invariances}

Just as in \cite{dt1,dt2,dt3}, there are residual gauge invariances.
Put another way, because under a gauge transformation,
$C \mapsto C + d B$, if two 1-gerbe connections $B$ differ by a 
1-gerbe gauge transformation, then they define the same action
on $C$.  Put another way still, only equivalence classes of 1-gerbes
define distinct gauge transformations of a $C$ field.
In addition, the maps between the 1-gerbes also have residual
gauge transformations.  For example, in the next section we shall
see that if the 1-gerbes $\Upsilon^g$ are topologically trivial,
then the maps $\Omega^{g_1, g_2}$ become bundles, defining gauge
transformations of the $\Upsilon^g$.  However, just as in
\cite{dt1,dt2,dt3}, only equivalence classes of bundles with connection
define distinct gauge transformations of a 1-gerbe, so we have a second
level of residual gauge invariances.

In the next two subsections we shall work out these two
classes of residual gauge invariances in more detail.

\subsubsection{First level of residual gauge invariances}

The first level of residual gauge invariances that we consider
consists of replacing one set of 1-gerbes with connection
(partially defining the difference between two orbifold group
actions) with an isomorphic set of 1-gerbes with connection.
In other words, since only equivalence classes of 1-gerbes with
connection act on a 2-gerbe with connection, how does the data
defining the difference between two orbifold group actions
change when the 1-gerbes are replaced with isomorphic 1-gerbes?

Suppose $\left( \kappa^g, \chi(g)^{\alpha} \right)$
defines a map between two isomorphic 1-gerbes with
connection:
\begin{displaymath}
\left( \kappa^g, \chi(g)^{\alpha} \right): \:
\left( \Upsilon^g, {\cal B}(g)^{\alpha}, {\cal A}(g)^{\alpha \beta} \right)
\: \longrightarrow \:
\left( \overline{\Upsilon}^g, \overline{ {\cal B} }(g)^{\alpha},
\overline{ {\cal A} }(g)^{\alpha \beta} \right)
\end{displaymath}
In other words, at the level of \v{C}ech cochain data, suppose
that the data defining the two sets of 1-gerbes with connection
is related as follows:
\begin{eqnarray*}
\overline{ \Upsilon }^g_{\alpha \beta \gamma} & = &
\left( \Upsilon^g_{\alpha \beta \gamma} \right) \,
\left( \kappa^g_{\alpha \beta} \right) \,
\left( \kappa^g_{\beta \gamma} \right) \,
\left( \kappa^g_{\gamma \alpha} \right) \\
\overline{ {\cal B}}(g)^{\alpha} & = &
{\cal B}(g)^{\alpha} \: - \: d \chi(g)^{\alpha} \\
\overline{ {\cal A}}(g)^{\alpha \beta} & = &
{\cal A}(g)^{\alpha \beta} \: - \: \chi(g)^{\alpha} \: + \:
\chi(g)^{\beta} \: + \: d \log \kappa^g_{\alpha \beta}
\end{eqnarray*}

Recall that in 
order to define an orbifold group action, in addition to specifying
1-gerbes with connection, we also must specify maps 
$\left( \Omega^{g_1, g_2}, \theta(g_1, g_2) \right)$ between
the 1-gerbes with connection, as well as maps $\omega^{g_1, g_2, g_3}$
between the maps between the maps between the 1-gerbes with connection.
If we replace one set of 1-gerbes with connection with an isomorphic
set of 1-gerbes with connection as above, then it is straightforward
to check that the remaining data defining the difference between
orbifold group actions changes as follows:
\begin{eqnarray*}
\overline{\Omega}^{g_1, g_2}_{\alpha \beta} & = &
\left( \Omega^{g_1, g_2}_{\alpha \beta} \right) \,
\left( \kappa^{g_1 g_2}_{\alpha \beta} \right) \,
\left( \kappa^{g_2}_{\alpha \beta} \right)^{-1} \,
\left( g_2^* \kappa^{g_1}_{\alpha \beta} \right)^{-1} \\
\overline{\theta}(g_1, g_2)^{\alpha} & = &
\theta(g_1, g_2)^{\alpha} \: + \: \chi(g_1 g_2)^{\alpha}
\: - \: \chi(g_2)^{\alpha} \: - \: g_2^* \chi(g_1)^{\alpha} \\
\overline{\omega}^{g_1, g_2, g_3}_{\alpha} & = &
\omega^{g_1, g_2, g_3}_{\alpha}
\end{eqnarray*}
In short, the maps $\left( \Omega^{g_1, g_2}, \theta(g_1, g_2) \right)$
are changed when the 1-gerbes are replaced with isomorphic 1-gerbes
with connection, but the maps between maps ($\omega$) remain invariant.

\subsubsection{Second level of residual gauge invariances}
\label{secondresid}

Even if we hold fixed the 1-gerbes with connection
$\left( \Upsilon^g, {\cal B}(g), {\cal A}(g) \right)$,
we still have a remaining residual gauge invariance,
produced by replacing the maps $\left( \Omega^{g_1, g_2}, \theta(g_1, g_2) 
\right)$ with isomorphic maps.

Let $\kappa^{g_1, g_2}$ define a map between two such maps,
as
\begin{displaymath}
\kappa^{g_1, g_2}: \: \left( \Omega^{g_1, g_2}, \theta(g_1, g_2) \right)
\: \longrightarrow \: \left( \overline{\Omega}^{g_1, g_2}, 
\overline{\theta}(g_1, g_2) \right)
\end{displaymath}
At the level of \v{C}ech cochains we can describe this map as
\begin{eqnarray*}
\overline{\Omega}^{g_1, g_2}_{\alpha \beta} & = &
\left( \Omega^{g_1, g_2}_{\alpha \beta} \right) \,
\left( \kappa^{g_1, g_2}_{\alpha} \right) \,
\left( \kappa^{g_1, g_2}_{\beta} \right)^{-1} \\
\overline{\theta}(g_1, g_2)^{\alpha} & = &
\theta(g_1, g_2)^{\alpha} \: + \:
d \log \kappa^{g_1, g_2}_{\alpha}
\end{eqnarray*}

It is straightforward to check that if we change
$\left( \Omega^{g_1, g_2}, \theta(g_1, g_2) \right)$ as above,
then we must also alter the maps between maps as
\begin{displaymath}
\overline{\omega}^{g_1, g_2, g_3}_{\alpha} \: = \:
\left( \omega^{g_1, g_2, g_3}_{\alpha} \right) \,
\left( \kappa^{g_1 g_2, g_3}_{\alpha} \right) \,
\left( g_3^* \kappa^{g_1, g_2}_{\alpha} \right) \,
\left( \kappa^{g_1, g_2 g_3}_{\alpha} \right)^{-1} \,
\left( \kappa^{g_2, g_3}_{\alpha} \right)^{-1}
\end{displaymath}

Formally, if both the $\omega$ and $\kappa$ were constant maps,
note that the equation above would amount to the statement that
$\overline{\omega}$ and $\omega$ differ by a group 3-coboundary
defined by $\kappa$.

\subsection{$H^3(\Gamma, U(1))$}

At least part (and sometimes all) of the possible differences
between orbifold group actions are described by elements of
the group cohomology group\footnote{With trivial action on the
coefficients.} $H^3(\Gamma, U(1))$, as we shall now describe.

Recall from \cite{dt2,dt3} that in order to find $H^2(\Gamma, U(1))$
in data consisting of a set of bundles $T^g$ with connection,
together with
isomorphisms $\omega^{g,h}$, one took the bundles to be topologically
trivial, and the connections on those bundles to be gauge-trivial.
The resulting degrees of freedom resided essentially entirely in
the isomorphisms $\omega^{g,h}$, and were counted by $H^2(\Gamma, U(1))$.
Also, because the bundles were trivial, and the gauge-connections
were trivial, Wilson surfaces $\exp \left( \int B \right)$
around Riemann surfaces created by the orbifold group action
had a constant phase, independent of the details of the Riemann
surface, depending only on the orbifold group elements used to glue together
the sides.

We find $H^3(\Gamma, U(1))$ in the present case in a similarly
restrictive context.
Take the 1-gerbes $\Upsilon^g$ to be canonically trivial
(meaning, all transition functions identically 1),
and assume all the ${\cal A}(g)^{\alpha \beta} = 0$,
so the $B$ field is a globally-defined 2-form.
(Equivalently, take the 1-gerbes $\Upsilon^g$ to be trivial
with gauge-trivial connection; then we can map to the situation
just described to get an equivalent action on the $C$ field.)

In this case, the 1-gerbe maps $\Omega^{g_1, g_2}$
become principal $U(1)$ bundles, from equation~(\ref{Omasmap}).
(This is just another way of saying that a 1-gerbe map from a gerbe
into itself is a gauge transformation of the 1-gerbe, and such
gauge transformations are defined by bundles.)
Also, from equation~(\ref{calAthetaOm})
we see that $\theta(g_1, g_2)^{\alpha}$ define a connection
on the principal $U(1)$ bundle $\Omega^{g_1, g_2}$.
Also, from equation~(\ref{thetaom}) we see that the bundle morphism
$\omega^{g_1, g_2, g_3}$ preserves the connection on the bundles.

Next, in order to find $H^3(\Gamma, U(1))$, we need to make
another restriction.  Take the bundles $\Omega^{g_1, g_2}$ to
all be topologically trivial, and the connections
$\theta(g_1, g_2)$ to all be gauge-trivial.

Again, only equivalence classes of bundles with connection
are meaningful, so this data can be equivalently mapped to
the bundles $\Omega^{g_1, g_2}$ being canonically trivial and
the connections $\theta(g_1, g_2)$ identically zero.
We see from equation~(\ref{thetaom}) that the 
maps $\omega^{g_1, g_2, g_3}$ must be constant.

Assuming the covering space is connected, the maps $\omega^{g_1, g_2, g_3}$
define a map $\Gamma \times \Gamma \times \Gamma \rightarrow U(1)$.
Also, from equation~(\ref{xiconstr}) we see that these maps
satisfy the group 3-cocycle condition.
Finally, recall that we still have a residual gauge invariance:
we can gauge-transform the bundles $\Omega^{g_1, g_2}$ by a 
constant gauge transformation,
as described in section~\ref{secondresid}.  It is easy to see that such
gauge transformations change the maps $\omega^{g_1, g_2, g_3}$
by a group 3-coboundary. 

Thus, we find that (many) differences between orbifold group
actions are classified by elements of the group
cohomology group $H^3(\Gamma, U(1))$, where the group 3-cocycles
ultimately come from the maps $\omega^{g_1, g_2, g_3}$.

As a check, it is straightforward to show that the
group $H^3(\Gamma, U(1))$ also enters cohomology calculations
in simple cases.  
As a simple example, consider possible holonomies of a 
3-form potential on an $n$-torus $T^n$.
Construct $T^n$ as a quotient:  $T^n = {\bf R}^n / {\bf Z}^n$.
Now, \cite[section III.1]{brown}
\begin{displaymath}
H_i({\bf Z}, {\bf Z}) \: = \: \left\{ \begin{array}{cl}
                                   {\bf Z} & i = 0, 1 \\
                                   0    & i > 1
                                   \end{array} \right.
\end{displaymath}
so from the K\"unneth formula \cite[section V.5]{brown} we find that
\begin{displaymath}
H_3( {\bf Z}^n, {\bf Z}) \: = \: {\bf Z}^k
\end{displaymath}
where
\begin{displaymath}
k \: = \:
\left( \begin{array}{c} n \\ 3 \end{array} \right)
\end{displaymath}
and from the universal coefficient theorem
\cite[section III.1 ex. 3]{brown}
\begin{eqnarray*}
H^3({\bf Z}^n, U(1)) & = & \mbox{Hom}\left( H_3( {\bf Z}^n, {\bf Z} ),
U(1) \right) \\
& = & U(1)^k
\end{eqnarray*}
so we find that if we orbifold the trivial $C$ field (i.e.,
$C \equiv 0$) on ${\bf R}^n$ by the freely-acting ${\bf Z}^n$,
the number of possible ways to combine the action of the orbifold
group with a gauge transformation of the $C$ fields is counted by
\begin{displaymath}
 \left[ U(1)
\right]^k
\end{displaymath}
which happily coincides with the possible holonomies of flat
(and topologically trivial) $C$ fields on $T^n$.
Thus, we should not be surprised to find the group
$H^3(\Gamma, U(1))$ appearing in our calculation of analogues
of discrete torsion for the $C$ field.

\subsection{Detailed classification of orbifold group actions}

In the last subsection we described how the differences between
many orbifold group actions were described by elements of $H^3(\Gamma, U(1))$.
Under what circumstances do those represent all of the differences,
and what do additional differences look like?

Let $X$ denote the covering space.
Suppose $H^3(X, {\bf Z})$ has no torsion,
$H^2(X, {\bf Z}) = 0$, and $\pi_1(X) = 0$.
Then all orbifold group actions on $C$ fields differ by an element
of $H^3(\Gamma, U(1))$.  This is straightforward to check.
Since $H^3(X, {\bf Z})$ contains no torsion, all the flat 1-gerbes
$\Upsilon^g$ must be topologically trivial, and so can be mapped
to the canonical trivial 1-gerbe.  Since $H^2(X, {\bf Z}) = 0$,
any connection on the 1-gerbes $\Upsilon^g$ must be gauge-trivial, 
and so we can
map them all to the zero connection, without loss of generality.
The 1-gerbe maps $(\Omega^{g_1, g_2}, \theta(g_1, g_2))$ are now
bundles with connection.  However, because $H^2(X, {\bf Z}) = 0$
and $\pi_1(X) = 0$, the bundles $\Omega^{g_1, g_2}$ must be topologically
trivial, and the connection $\theta(g_1, g_2)$ must be gauge-trivial.

More generally, however, one can expect to find additional possible
orbifold group actions, just as for $B$ fields in \cite{dt3},
{\it i.e.} analogues for $C$ fields of shift orbifolds \cite{dtshift}.
This is certainly consistent with naive examinations of homology.
From the Cartan-Leray spectral sequence \cite[section VII.7]{brown}:
\begin{displaymath}
E^2_{p,q} \: = \: H_p \left( \Gamma, H_q(X, {\bf Z}) \right)
\: \Longrightarrow \: H_{p+q}(X/\Gamma, {\bf Z})
\end{displaymath}
Ignoring differentials for a moment, we can see immediately
that the homology $H_3(X/\Gamma, {\bf Z})$ involves more than
just $H_3(X, {\bf Z})$ and $H_3(\Gamma, {\bf Z})$
(which dualizes to $H^3(\Gamma, U(1))$; there can also
be nontrivial contributions involving $H_1(X, {\bf Z})$ and
$H_2(X, {\bf Z})$.
 
In the remainder of this article, we shall concentrate
on the differences between orbifold group actions classified
by $H^3(\Gamma, U(1))$, and will largely ignore the additional
possible orbifold group actions.  We will return to these
additional degrees of freedom in \cite{dtshift}, where we shall
argue that they encode the M-theory dual of IIA discrete torsion.

\subsection{Commentary}

In this section, we have studied orbifold group actions
on $C$ fields, and shown, for example, that the group
$H^3(\Gamma, U(1))$ arises in describing the differences
between (some) orbifold group actions.

We should point out that our analysis does not depend
upon whether or not the orbifold group $\Gamma$ acts freely,
just as our analysis of $B$ fields in \cite{dt1,dt2,dt3}
did not depend upon whether $\Gamma$ acts freely.
(Now, understanding M theory on singular spaces can be somewhat
subtle, so there may be physics subtleties; however, our mathematical
analysis is independent of whether or not $\Gamma$ acts freely.)

We should also point out that our analysis does not
depend upon whether or not $\Gamma$ is abelian,
just as our analysis of $B$ fields in \cite{dt1,dt2,dt3}
did not depend upon whether $\Gamma$ is abelian.

We should also mention that our analysis does
not depend upon the curvature of the $C$ field being zero
in integral cohomology $H^4({\bf Z})$, just as for our
analysis of $B$ fields in \cite{dt1,dt2,dt3}.
However, if our 2-gerbe is topologically nontrivial,
then we need to check that an action of the orbifold group
exists -- just as for vector fields and $B$ fields,
topologically nontrivial objects do not always admit
orbifold group actions.  The analysis we have provided
continues to hold on the assumption that orbifold group actions exist.

\section{Membrane twisted sector phases}   \label{twphase}

Discrete torsion (for $B$ fields) was originally
discovered in \cite{vafa1} as a phase ambiguity in
twisted sector contributions to string loop partition functions.
As noted in \cite{dt1,dt3},
this phase ambiguity arises because of the presence of
a term $\int B$ in the worldsheet sigma model.

Now, low-energy effective actions for membrane worldvolumes
contain a term $\int C$, so one would expect that
``membrane twisted sectors'' would be weighted by factors
derived from $\exp \left( \int B \right)$.

In this section we shall derive 
these membrane worldvolume factors, the analogue
of the twisted sector phases of \cite{vafa1} for membranes.
We shall also check that these factors are invariant
under the natural $SL(3,{\bf Z})$ action on $T^3$.

At this point we should emphasize that we do not wish to imply that
we believe that M theory is a theory of membranes; rather, we are merely
computing how the low-energy effective action on a membrane worldvolume
sees our degrees of freedom.  We should also reiterate that our
analysis is too naive physically in that we have ignored
gravitational corrections and underlying $E_8$ structures in the
M-theory $C$ form \cite{edgreg1,edgreg2}, 
and instead blindly treated the $C$ form as a
connection on a 2-gerbe.

\subsection{Twisted sector phases on $T^3$}

For simplicity, we will assume that $C \equiv 0$
(and that the 2-gerbe is topologically trivial), so that
any orbifold group action on the $C$ field can be described
solely in terms of its difference from the identity.

Furthermore, for simplicity we shall also only consider those
orbifold group actions described by elements of $H^3(\Gamma, U(1))$.
This means that we shall take the 1-gerbes $\Upsilon^g$ to all
be canonically topologically trivial, with flat ${\cal B}(g)^{\alpha}$
and, for additional simplicity, ${\cal A}(g)^{\alpha \beta}
\equiv 0$.  The gerbe maps $\left( \Omega^{g_1, g_2}, \theta(g_1, g_2) 
\right)$ can now be interpreted as principal $U(1)$ bundles with
connection.  We assume each bundle $\Omega^{g_1, g_2}$ is topologically
trivial, and that the connection $\theta(g_1, g_2)$ is gauge-trivial.

\begin{figure}
\centerline{\psfig{file=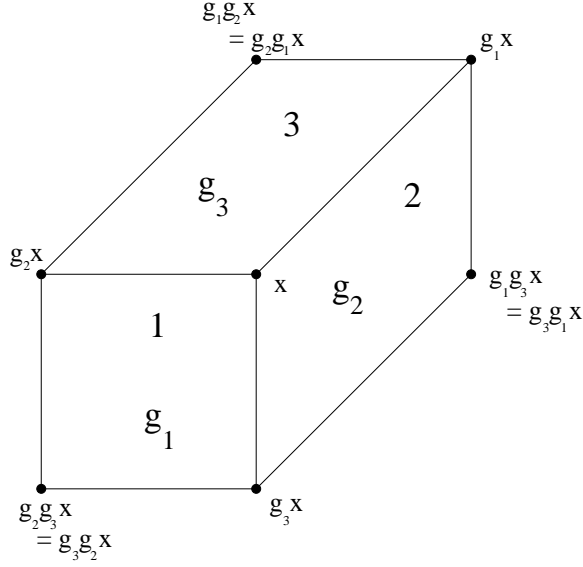,width=3in}}
\caption{ \label{figcbox}
Three-torus seen as open box on covering space.}
\end{figure}

Consider a $T^3$ in the quotient space, which looks like a box
in the covering space with sides identified by the orbifold group
action, as illustrated in figure~(\ref{figcbox}). 

Following the procedure described in \cite{dt3},
from glueing the faces of the cube together, one would
naively believe that the holonomy
\begin{displaymath}
\exp \, \left( \, \int C \, \right)
\end{displaymath}
over the volume of the cube on the covering space
should be corrected by a phase factor
\begin{equation}    \label{firstnaive}
\exp \, \left( \, \int_1 {\cal B}(g_1) \: + \:
\int_2 {\cal B}(g_2) \: + \: \int_3 {\cal B}(g_3) \, \right)
\end{equation}

However, just as in \cite{dt3}, this can not be the complete
answer, if for no other reason than it is not invariant under
gauge transformations of the 1-gerbes.  We can go a long way towards
fixing this difficulty by closely examining the edges at the borders
of each face.

\begin{figure}
\centerline{\psfig{file=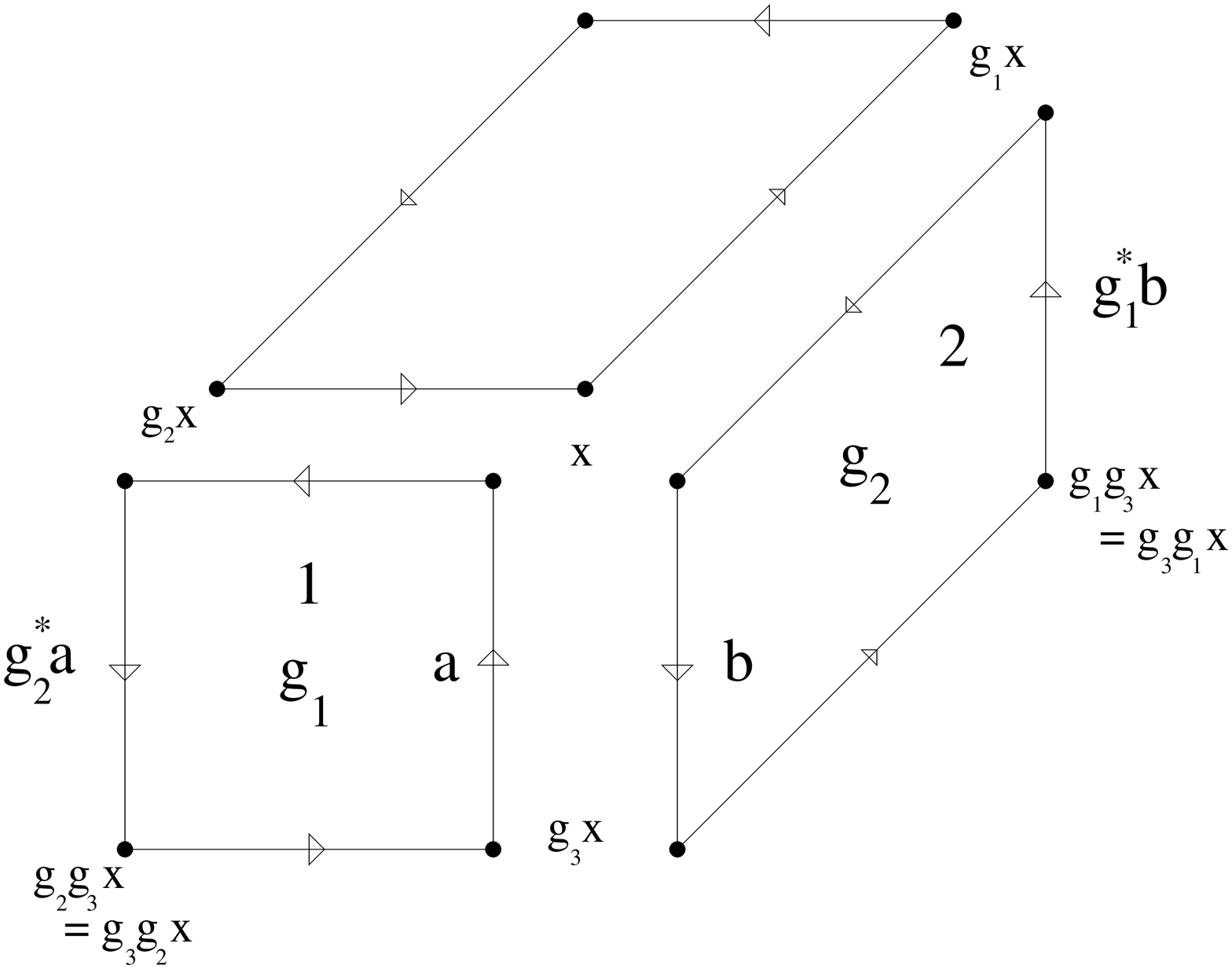,width=3in}}
\caption{ \label{figcbox2a}
Four of the twelve edges, all descending to same line.}
\end{figure}

Consider the edges shown in figure~(\ref{figcbox2a}),
labelled $a$, $b$, $g_2^* a$, and $g_1^* b$,
between faces 1 and 2.  These edges are all mapped to the same
curve in the quotient space by the orbifold group action.
From the expression
\begin{displaymath}
{\cal B}(g_1 g_2)^{\alpha} \: = \: {\cal B}(g_2)^{\alpha} \: + \:
g_2^* {\cal B}(g_1)^{\alpha} \: - \: d \theta(g_1, g_2)^{\alpha}
\end{displaymath}
we can derive that
\begin{displaymath}
\left[ \, {\cal B}(g_2) \: + \: g_2^* {\cal B}(g_1) \, \right]
\: - \: \left[ \, {\cal B}(g_1) \: + \: g_1^* {\cal B}(g_2) \, \right]
\: = \: d \left( \, \theta(g_1, g_2) \: - \: \theta(g_2, g_1) \, \right)
\end{displaymath}
so we can partially fix the naive expression~(\ref{firstnaive})
by adding the factor
\begin{equation}
\exp \, \left( - \int^{g_3 x}_x \left[ \, \theta(g_1, g_2) \: - \:
\theta(g_2, g_1) \, \right] \right)
\end{equation}
to take into account these four edges.

\begin{figure}
\centerline{\psfig{file=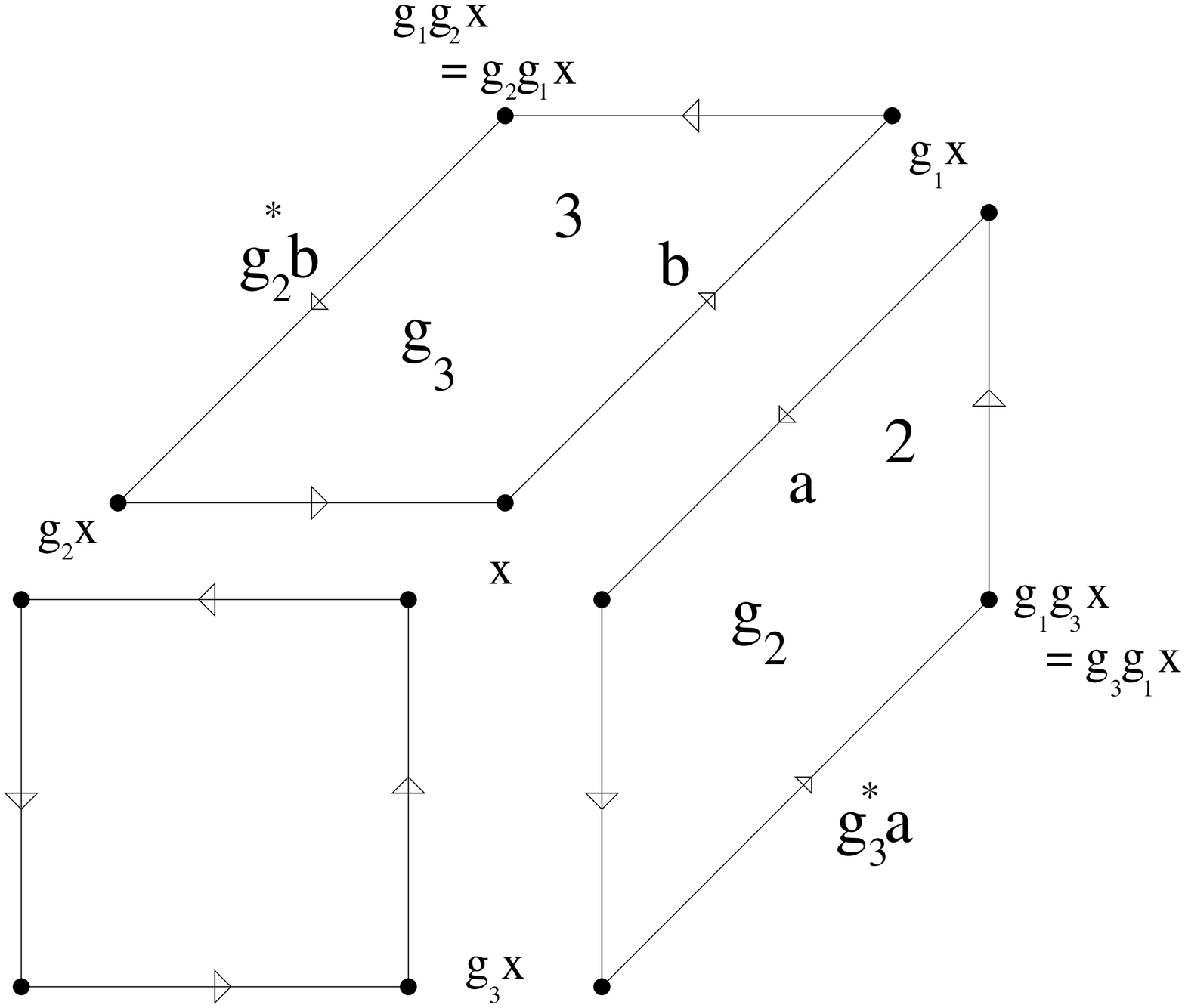,width=3in}}
\caption{ \label{figcbox2b}
Four of the twelve edges, all descending to same line.}
\end{figure}

Next, consider the edges shown in figure~(\ref{figcbox2b}),
labelled as before, between faces 2 and 3.
Proceeding as above, we find that these edges contribute
a factor to expression~(\ref{firstnaive}) given by
\begin{equation}
\exp \, \left( - \int^{g_1 x}_x \left[ \, \theta(g_2, g_3) \: - \:
\theta(g_3, g_2) \, \right] \right)
\end{equation}

\begin{figure}
\centerline{\psfig{file=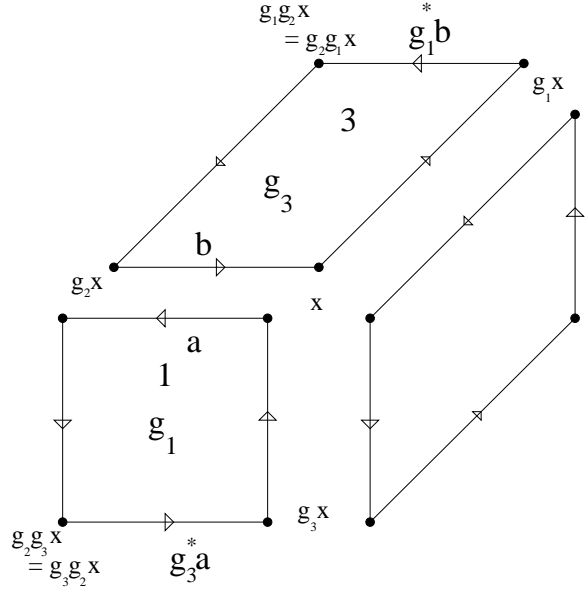,width=3in}}
\caption{ \label{figcbox2c}
Four of the twelve edges, all descending to same line.}
\end{figure}

Finally, consider the remaining four edges shown in figure~(\ref{figcbox2c}),
labelled as before, between faces 1 and 3.
Proceeding as above, we find that these edges contribute a factor
to expression~(\ref{firstnaive}) given by
\begin{equation}
\exp \, \left( - \int^{x}_{ g_2 x} \left[ \, \theta(g_1, g_3) \: - \:
\theta(g_3, g_1) \, \right] \right)
\end{equation}

To summarize our results so far, we have found a phase factor
(induced by gauge transformations at the boundaries) given by
\begin{eqnarray*}
\lefteqn{ \exp \, \left( \, - \: \int^{g_3 x}_x \left[ \, 
\theta(g_1, g_2) \: - \: \theta(g_2, g_1) \, \right] \: - \:
\int^{g_1 x}_x \, \left[ \, \theta(g_2, g_3) \: - \: \theta(g_3, g_2) \, 
\right] \, \right) } \\
\: & \: \: & \: \cdot \exp \, \left(
\: - \:
\int^x_{g_2 x} \, \left[ \, \theta(g_1, g_3) \: - \: \theta(g_3, g_1) \, 
\right]
\, \right)  \:
% \: & \: \: & \: 
\cdot \: \exp \, \left(
\, \int_1 {\cal B}(g_1) \: + \:
\int_2 {\cal B}(g_2) \: + \: \int_3 {\cal B}(g_3) \, \right)
\end{eqnarray*}

However, the phase factor above can still not be completely
correct, because it is not invariant under gauge transformations
of the bundles $\Omega^{g_1, g_2}$.  We need to add an additional factor
to fix the contributions from corners of the box.

In order to determine which additional factor to add,
note that the connections $\theta$ at the ends of the line integrals
(i.e., the corners of the box) are given by
\begin{eqnarray*}
\lefteqn{ g_3^* \theta(g_1, g_2) \: - \: g_3^* \theta(g_2, g_1)
\: - \: \theta(g_1, g_2) \: + \: \theta(g_2, g_1) } \\
 \: & \: \: & \: + \: g_1^* \theta(g_2, g_3) \: - \:
g_1^* \theta(g_3, g_2) \: - \: \theta(g_2, g_3) \: + \:
\theta(g_3, g_2) \\
 \: & \: \: & \: - \: g_2^* \theta(g_1, g_3) \: + \:
g_2^* \theta(g_3, g_1) \: + \: \theta(g_1, g_3) \: - \: \theta(g_3, g_1)
\end{eqnarray*}
Using the identity
\begin{displaymath}
\theta(g_2, g_3) \: + \: \theta(g_1, g_2 g_3) \: = \:
g_3^* \theta(g_1, g_2) \: + \: \theta(g_1 g_2, g_3) \: - \:
d \log \omega^{g_1, g_2, g_3}
\end{displaymath}
it is straightforward to check that the twelve-term corner contribution
sum above is equal to
\begin{displaymath}
d \, \left[ \, \log \omega^{g_1, g_2, g_3} \: - \:
\log \omega^{g_2, g_1, g_3} \: - \: \log \omega^{g_3, g_2, g_1} \: + \:
\log \omega^{g_3, g_1, g_2} \: + \: \log \omega^{g_2, g_3, g_1}
\: - \: \log \omega^{g_1, g_3, g_2} \, \right]
\end{displaymath}

Using this fact, we can now write the correct, completely gauge-invariant,
phase factor picked up by $\exp \left( \int C \right)$
because of gauge transformations at boundaries:
\begin{eqnarray*}
\lefteqn{ \left( \omega^{g_1, g_2, g_3}_x \right) \,
\left( \omega^{g_2, g_1, g_3}_x \right)^{-1} \,
\left( \omega^{g_3, g_2, g_1}_x \right)^{-1} \,
\left( \omega^{g_3, g_1, g_2}_x \right) \,
\left( \omega^{g_2, g_3, g_1}_x \right) \,
\left( \omega^{g_1, g_3, g_2}_x \right)^{-1} } \\
\: & \: \: & \: \cdot \exp \, \left( \, - \: \int^{g_3 x}_x \left[ \, 
\theta(g_1, g_2) \: - \: \theta(g_2, g_1) \, \right] \: - \:
\int^{g_1 x}_x \, \left[ \, \theta(g_2, g_3) \: - \: \theta(g_3, g_2) \, 
\right] \, \right) \\
\: & \: \: & \: \cdot \exp \, \left( 
\: - \:
\int^x_{g_2 x} \, \left[ \, \theta(g_1, g_3) \: - \: \theta(g_3, g_1) \, 
\right]
\, \right)  \\
 \: & \: \: & \: \cdot \exp \, \left(
\, \int_1 {\cal B}(g_1) \: + \:
\int_2 {\cal B}(g_2) \: + \: \int_3 {\cal B}(g_3) \, \right)
\end{eqnarray*}

Since the expression above is gauge-invariant, we can evaluate
it by evaluating it in any convenient gauge.
Consider the gauge in which ${\cal B}(g) \equiv 0$ for all $g$,
and $\theta(g_1, g_2) \equiv 0$ for all $g_1$, $g_2$.
Then the $\omega^{g_1, g_2, g_3}$ become constant maps into $U(1)$
(assuming the covering space is connected, of course),
describing a group 3-cocycle.  In this gauge, it is manifest that
the expression above is equal to
\begin{equation}    \label{t3phase}
\left( \omega^{g_1, g_2, g_3} \right) \,
\left( \omega^{g_2, g_1, g_3} \right)^{-1} \,
\left( \omega^{g_3, g_2, g_1} \right)^{-1} \,
\left( \omega^{g_3, g_1, g_2} \right) \,
\left( \omega^{g_2, g_3, g_1} \right) \,
\left( \omega^{g_1, g_3, g_2} \right)^{-1}
\end{equation}

As a check, note that expression~(\ref{t3phase}) is invariant
under changing the group 3-cocycles by coboundaries,
as indeed it must be in order to give associate a well-defined
phase to elements of $H^3(\Gamma, U(1))$.

\subsection{``Modular invariance'' on $T^3$}

The twisted sector phases originally defining discrete torsion in \cite{vafa1}
had the property that they were invariant under the action of the modular group
$SL(2, {\bf Z})$ of $T^2$.
In this section we shall review that calculation for
$B$ field discrete torsion, and then check that the phases
we just computed for ``membrane twisted sectors'' are invariant
under the natural $SL(3,{\bf Z})$ acting on $T^3$.

We do not wish to imply that there necessarily exists a precise notion of
modular invariance for membranes.  Rather, we are merely observing that
the $SL(2, {\bf Z})$ invariance of the twisted sector phases in
\cite{vafa1} has an analogue here, that the ``twisted sector phases''
seen by the low-energy effective action on a membrane worldvolume
are $SL(3, {\bf Z})$ invariant.  In particular, we have derived these
membrane phases from more fundamental considerations, and at the end
of the day, we are observing an $SL(3, {\bf Z})$ invariance. 
In other words,
unlike the original discrete torsion story of \cite{vafa1},
we are not using $SL(3, {\bf Z})$ invariance as a starting point,
but rather noting it as a consequence.

\subsubsection{Review of modular invariance on $T^2$}

Before checking that the membrane twisted sector phases for
$T^3$ are invariant under $SL(3,{\bf Z})$, we shall first take a moment
to review how this works for standard discrete torsion and twisted
sector phases on $T^2$.

Recall that in a genus one partition function, 
the phase $\epsilon(g,h)$ associated to a twisted sector
determined by the commuting pair $(g,h)$ is given by
\begin{displaymath}
\epsilon(g,h) \: = \: \left( \omega^{g,h} \right) \,
\left( \omega^{h,g} \right)^{-1}
\end{displaymath}
where $\omega^{g,h}$ is the group 2-cocycle describing an element
of $H^2(\Gamma, U(1))$.

An element of $SL(2,{\bf Z})$ described by the matrix
\begin{displaymath}
{\bf A} \: = \: \left( \begin{array}{cc}
        a & b \\
        c & d
       \end{array}
\right)
\end{displaymath}
maps
\begin{eqnarray*}
g & \mapsto & g^a \, h^b \\
h & \mapsto & g^c \, h^d 
\end{eqnarray*}

In terms of twisted sector phases, modular invariance
is the constraint that
\begin{equation} \label{t2modinv}
\epsilon\left( \, g^a h^b, \, g^c h^d \, \right) \: = \:
\epsilon(g,h)
\end{equation}

To prove equation~(\ref{t2modinv}), we use the following identities:
\begin{enumerate}
\item $\epsilon(g,h) = \epsilon(h,g)^{-1}$  (by inspection) 
\item $\epsilon(g_1 g_2, g_3) \: = \: \epsilon(g_1, g_3) \, 
\epsilon(g_2, g_3)$  
\end{enumerate}
(The second identity is proven by manipulating the group cocycle
condition and doing some algebra.)

From these identities, it is straightforward to verify
\begin{eqnarray*}
\epsilon\left( \, g^a h^b, \, g^c h^d \, \right)
& = & \epsilon(g,h)^{\det {\bf A}} \\
& = & \epsilon(g,h)
\end{eqnarray*}
and so modular invariance is proven.

\subsubsection{Calculation on $T^3$}

Consider a twisted sector determined by the three commuting
elements $(g_1, g_2, g_3)$.
Under the action of an element ${\bf A}$ of $SL(3,{\bf Z})$ given by
\begin{displaymath}
{\bf A} \: = \: \left( \begin{array}{ccc}
                       a_{11} & a_{12} & a_{13} \\
                       a_{21} & a_{22} & a_{23} \\
                       a_{31} & a_{32} & a_{33} 
                       \end{array}
                       \right)
\end{displaymath}
the orbifold group elements $(g_1, g_2, g_3)$ transform as
\begin{eqnarray*}
g_1 & \mapsto & g_1^{a_{11}} \, g_2^{a_{12}} \, g_3^{a_{13}} \\
g_2 & \mapsto & g_1^{a_{21}} \, g_2^{a_{22}} \, g_3^{a_{23}} \\
g_3 & \mapsto & g_1^{a_{31}} \, g_2^{a_{32}} \, g_3^{a_{33}}
\end{eqnarray*}

Define $\epsilon(g_1, g_2, g_3)$ to be the twisted sector phase
associated to the group 3-cocycle $\omega^{g_1, g_2, g_3}$:
\begin{displaymath}
\epsilon(g_1, g_2, g_3) \: = \:
\left( \omega^{g_1, g_2, g_3} \right) \,
\left( \omega^{g_2, g_1, g_3} \right)^{-1} \,
\left( \omega^{g_3, g_2, g_1} \right)^{-1} \,
\left( \omega^{g_3, g_1, g_2} \right) \,
\left( \omega^{g_2, g_3, g_1} \right) \,
\left( \omega^{g_1, g_3, g_2} \right)^{-1}
\end{displaymath}

In order for ``modular invariance'' to hold, i.e., in order for
the twisted sector phases to be invariant under $SL(3,{\bf Z})$,
it had better be the case that
\begin{equation}    \label{t3modinv}
\epsilon\left( \, g_1^{a_{11}}  \, g_2^{a_{12}}  \, g_3^{a_{13}} , \:
g_1^{a_{21}}  \, g_2^{a_{22}}  \, g_3^{a_{23}} , \: 
 g_1^{a_{31}} \, g_2^{a_{32}} \, g_3^{a_{33}} \, \right)
\: = \: \epsilon(g_1, g_2, g_3)
\end{equation}

In order to verify equation~(\ref{t3modinv}),
it is useful to first note the following identities:
\begin{enumerate}
\item $\epsilon(g_1, g_2, g_3)$ is antisymmetric under interchange
of any two of the group elements -- e.g., $\epsilon(g_1, g_2, g_3) 
= \epsilon(g_2, g_1, g_3)^{-1}$
\item $\epsilon(g_1 g_2, g_3, g_4) = \epsilon(g_1, g_3, g_4) \,
\epsilon(g_2, g_3, g_4)$
\end{enumerate}
The first identity is trivial to verify; the second requires a significant
amount of algebraic manipulation of the 3-cocycle condition.

Given these identities, it is straightforward to verify that
\begin{eqnarray*}
\epsilon\left( \, g_1^{a_{11}}  \, g_2^{a_{12}}  \, g_3^{a_{13}} , \:
g_1^{a_{21}}  \, g_2^{a_{22}}  \, g_3^{a_{23}} , \:
 g_1^{a_{31}} \, g_2^{a_{32}} \, g_3^{a_{33}} \, \right)
& = & \epsilon(g_1, g_2, g_3)^{ \det {\bf A}} \\
& = & \epsilon(g_1, g_2, g_3)
\end{eqnarray*}
and so we see that ``modular invariance'' holds for membranes on $T^3$.

Again, we do not wish to imply that there necessarily exists a precise
physical analogue of modular invariance for membranes.
Rather, having calculated the analogues of twisted sector phases
from more fundamental considerations, we are merely noting that those
phases are $SL(3, {\bf Z})$ invariant, a precise analogue of the
modular invariance constraint used in \cite{vafa1} to
derive the original $B$ field discrete torsion.

\section{Notes on local orbifolds}

In \cite{dt3} we spoke very briefly about the possibility
of local orbifold degrees of freedom for $B$ field discrete
torsion.  To review, given a quotient space such as
$T^4/{\bf Z}_2$, for example, in addition to degrees of freedom
obtained by describing the space as a global orbifold, there can
sometimes be additional degrees of freedom.  For example,
from the discussion in \cite{dt1,dt3}, there are 2 possible
${\bf Z}_2$ orbifold group actions on a trivial line bundle
on $T^4$.  However, if we view $T^4/{\bf Z}_2$ as a collection
of coordinate patches of the form ${\bf C}^2/{\bf Z}_2$,
each patch has a 2-element degree of freedom obtained from
possible ${\bf Z}_2$ actions on a trivial line bundle on ${\bf C}^2$,
and when we glue the patches back together, we find more consistent
degrees of freedom than just the 2 global ones.
The additional degrees of freedom come from describing the orbifold
group action on the line bundle locally, rather than globally,
and so we refer to them as local orbifold degrees of freedom.
A theory with such local orbifold degrees of freedom ``turned on''
cannot be obtained by a global orbifold, even though the underlying
space can be.

We do not claim to have studied possible local orbifold degrees of
freedom for analogues of discrete torsion for $C$ fields;
however, in this section we merely wish to point out their
possible existence.

In particular, readers wishing to describe all possible
compactifications of M theory to 7 dimensions, for example,
would need to understand not only global degrees of freedom
corresponding to analogues of discrete torsion for the M theory
three-form potential $C$, but also possible local orbifold
degrees of freedom.

\section{Conclusions}

In this paper we have described orbifold group actions on
$C$ fields, i.e., the analogue of discrete torsion for $C$ fields.
As pointed out in \cite{dt1,dt2,dt3}, discrete torsion (for $B$ fields)
is a purely mathematical consequence of understanding orbifold
group actions on $B$ fields, and once one understands that
orbifold group action on $B$ fields, it is straightforward
to work out the correct version for $C$ fields.

As mentioned in the introduction, it would be naive to immediately
read physical results into our calculation.  We have treated
the $C$ field in isolation, but in fact that is
certainly naive \cite{fluxquant}, and one should take
into account interaction terms in the theory which we have
neglected.

However, it is still very important to emphasize that in principle,
an analogue of discrete torsion exists for the M-theory three-form
potential $C$ and the other tensor field potentials of string
theory.  Although the calculation we have presented is naive,
it should make the point that some degrees of freedom
analogous to discrete torsion exist, and should be studied
in greater detail.

\section{Acknowledgements}

We would like to thank P.~Aspinwall, A.~Knutson, D.~Morrison,
and R.~Plesser for useful conversations.

\end{document}